# On the accurate calculation of the dielectric constant and the diffusion coefficient from molecular dynamics simulations: the case of SPC/E water

Orsolya Gereben and László Pusztai

*Research Institute for Solid State Physics and Optics, Hungarian Academy of Sciences, P.O.B 49, H-1525 Budapest, Hungary*

**Abstract**

The effect of the applied trajectory length on the convergence of the static dielectric constant and the self-diffusion coefficient were examined for the SPC/E water model in the NVT ensemble with different system size at 293 K. Very long simulation times of 6-8 ns were employed in order to track the convergence of these properties. Temperature dependence and isotope effects, via using $D_2O$ instead of $H_2O$, were also investigated. A simulation for the polarizable SWM4-DP model was also carried out to compare the effect of different potential models. Radial distribution functions and the neutron weighted structure factor were also calculated; they were found to be insensitive to changing the system size in the range of 216-16000 molecules. On the other hand, the static dielectric constant and the diffusion coefficient are rather sensitive to the applied trajectory length, system size and the method of calculation. These latter properties are therefore not appropriate for assessing, and distinguishing between, potential models of water. It is clearly shown that trajectories shorter than about 6 ns are not sufficient for a sufficiently accurate determination of the dielectric constant of this water model.

**1. Introduction**

Computer simulation of aqueous systems, including electrolyte and protein solutions, has been (and will increasingly be) an indispensable technique for the investigation of these systems. A key element of these computations is the water interaction potential model applied and therefore, precise knowledge concerning the behavior of the potentials, with respect to changing system size and simulation length, is of utmost importance.

Properties of the SPC/E[1,2,3,4,5] and SWM4-DP[6] water potential models have already been investigated thoroughly. The SPC/E model was chosen here, too, for its simplicity and popularity when studying aqueous systems, while the polarizable SWM4-DP model was selected as this model performed very well in our previous investigations regarding the structure of salt solutions[7,8]. As during these investigations we found, as side effects, quite substantial variances due to the application of different trajectory lengths while calculating the static dielectric constant, we decided to perform a new, systematic investigation on pure water. Our aim was not so much to give any further information on these specific models, but rather to exploit them as examples for studying the effects of different simulation parameters, mainly those of the applied trajectory length, system size and calculation method (this latter is only for the self-diffusion coefficient) during the determination of properties such as the static dielectric constant and the self-diffusion coefficient.

The dependence of the static dielectric constant on the system size using periodic boundary conditions, as well as reaction field, was previously investigated by Morrow[9]. He found that for a lattice of dipoles the static dielectric constant increases with increasing system size and slowly converges to the theoretical value of infinite system size under periodic boundary conditions. For the same lattice system, using a reaction field with $R_c$ cutoff distance, the convergence to the limit of infinite system size with increasing cutoff distance was not monotonous. In both cases large systems (~10000 dipoles for the periodic boundary condition model and $R_c$ larger than half of the system size) could produce dielectric constants within 10 % of the limiting value of infinite system size. For two small systems (125 and 343 molecules) Casulleras[10] found that the dielectric constant increases with increasing system size in molecular dynamics simulations performed for liquid methanol, using relatively short (500 ps) simulation lengths. It is fair to state that concerning the calculation of dielectric properties there are still quite a few ambiguities (see below) and thus, a systematic study by varying the system size and the length of the simulated trajectory (that is, the volume of configuration space explored) is rather timely.

The system-size dependence of the self-diffusion coefficient was studied earlier by Dünweg[11] for polymers, by Yeh[12] for TIP3P water and Lennard-Jones fluids under periodic boundary conditions and by Spångberg[13] for dilute (single ion) aqueous $Li^+$ and $Mg^{2+}$ solutions. An analytic correction based on hydrodynamic arguments proportional to $1/L$ (where $L$ is the simulation box size) was found. However, unfortunately, it is quite common in the water simulation literature that the finite-size effect is neglected when the self-diffusion coefficient is calculated. One of our aims therefore is to draw attention to the importance of this, apparently not unknown, finite-size effect.

We found that apart from the applied system size, the simulation trajectory length influences strongly the actual value of the self-diffusion constant. Another aim of ours was then to study the dependence of the self-diffusion coefficient on the calculation method applied (based on the mean square displacement or velocity auto correlation function) and the convergence of its value depending on the applied trajectory length and sampling interval.

The effects of temperature and isotopic substitution were also touched upon for one system size. Concerning the latter, it is known that classical molecular dynamics simulations (with the same interaction parameters are used, only the mass being different) can distinguish only between time-dependent properties of the isotopic compounds as the auto-correlation function, self-diffusion coefficient etc., but no difference is expected between equilibrium static properties such as the radial distribution function and static dielectric constant[14].

For studying the system-size effect of static structural properties, as well, the radial distribution function and the neutron weighted static structure factor were also calculated and compared.

## 2. Simulation details

The primary water potential in our investigations was the SPC/E model[1], at a temperature of 293 K. All simulations were carried out in the NVT ensemble under periodic boundary conditions with a number density of $\rho = 0.099$ Å$^{-3}$. Four different system sizes over a wide range were used, containing 216, 518, 2000 and 16000 water molecules; these systems are denoted by H1, H2, H3, H4, corresponding to the increasing number of molecules. To investigate the temperature dependence the H3-T model at 298 K was created, which calculation was otherwise identical to that on the H3 model. The isotope effect was examined by substituting hydrogen by deuterium. The D1 model is identical to H1 except that the mass of H atoms (1.00800 amu) changes to the mass of D atoms (2.014102 amu). Lennard-Jones (LJ) parameters for D$_2$O were the same as for H$_2$O and the same molecular geometry was assumed. Simulation H1-O was performed with 216 water molecules at 298 K, used slightly different run parameters applying long-range dispersion correction for the energy (as implemented in the molecular dynamics simulation package 'GROMACS') and shorter (1.25 fs) time step, as in this calculation we tried to reproduce the simulations performed by Svishchev and Kusalik[15] to make the comparison with the results of these authors more accurate. For a comparison with a different potential model, the GROMACS-adapted[16] version of the SWM4-DP potential was chosen (calculation name: 'SWM4-DP') which, in terms of system size and temperature, was identical to the H3 calculation. In our version of the SWM4-DP model the Drude particle was massless, no second thermostat for the oxygen-Drude pair was introduced and a time step of $dt$=2 fs was applied (see Table 1 for the summary of systems used).

All the molecular dynamics simulations were carried out by the GROMACS simulation package[17], using the leap-frog algorithm for integrating Newton's equations of motion with a time step $dt$=2 fs, if not stated otherwise. The main simulation parameters of the models are listed in Table 1. All simulations used the 6-12 LJ potential with a cutoff of 0.9 nm. The rigid molecular geometry was enforced with the SETTLE[18] algorithm. Calculations H1, H2, H3, H4, D1, H3-T and SWM4-DP were carried out with the same parameters (except for the differences explicitly given in Table 1) using particle-mesh Ewald[19] summation for the long range Coulomb-forces with interpolation order 4 and a grid size of 0.12 nm. The relative strength of the electrostatic interaction at the cutoff is ~$10^{-5}$, providing accuracy of $5 \cdot 10^{-3}$ which is still higher than the accuracy of the Lennard-Jones calculations. LJ parameters for SPC/E oxygen were $\sigma$ = 0.316557 nm, $\varepsilon$ = 0.650194 kJ/mol, and $\sigma$ = 0.31803 nm and $\varepsilon$ = 0.861141024 kJ/mol for the SWM4-DP model. Temperature was held constant with the help of a Berendsen thermostat[20] with temperature coupling time constant $\tau_T$ = 0.1 ps. The equilibration stage, the first 400 ps of the simulation, was discarded; data collection was carried out for the following 6-8.2 ns, as it is given in Table 1. Coordinates were collected at each time step for each system, apart from the largest H4 model, where every 10-th configuration was collected. (The size of the configuration file of already the mid-size 2000 molecule systems was around 300 Gbytes, which made the storage and handling of the file difficult.) Velocities for the calculation of the diffusion coefficient were collected for a much shorter period (see Table 1), restarting the simulation from the beginning of the data collection period.

### 3. Pair correlation functions and static structure factors

The average partial radial distribution functions, $g(r)$ were calculated for 4000 configurations (spanning 8000 ps in time) of the model systems, consecutive sample configurations being $\Delta t=2$ ps (or in case of H1-O, 1.25 ps) apart. As it can be seen in Figure 1a, there is virtually no difference regarding the partial $g(r)$'s. In Figure 1b the effect of slightly different parameters and temperature (H1, H1-O), and isotope substitution (D1) can be seen. There is no difference between H1, H1-O and D1 models, indicating that such small differences in terms of the parameters and the 5 K temperature difference did not have a visible effect on the $g(r)$. (Note that isotopic substitution using classical MD stimulation was not expected to show any structural change.) There is no visible difference between the corresponding $g(r)$ partials of the H3 and H3-T model, so the 5 K temperature increase for H3-T does not affect them (not shown).

On the other hand, there is a large difference between the size-wise and temperature-wise equal H3 and SWM4-DP models in terms of the radial distribution functions (Figure 2a), especially in the O-H and H-H partials. The neutron weighted total structure factor is very different, too (Figure 2b). This can be attributed partially to the fact that even the equilibrium geometry of the two models is different, in the SPC/E model the $d_{OH}$ =0.1 nm and the $\beta_{HOH}$ = 109.47 ° while in the SWM4-DP model the $d_{OH}$ =0.09572 nm and the $\beta_{HOH}$ = 104.52 °.

The effect of the trajectory length was also investigated for the small H1 model. The average $g(r)$ and $S(Q)$ were calculated for 10, 20, 50, 100, 1800 configurations (using the same time difference between consecutive configurations as before), apart from the previously used 4000 configuration calculation, corresponding to 20, 40, 100 and 3600 ps long trajectories. There was no difference either in the partial $g(r)$-s or in the partial and total $S(Q)$ for using 20-4000 configurations, and the result of the 10 configuration calculation hardly differed from them even for this small system size (not shown).

### 4. Static dielectric constant

During the calculation of the static dielectric constant (relative static permittivity), $\varepsilon$, $\varepsilon_{RF}=\infty$ was assumed. $\varepsilon$ was calculated for the SPC/E model according to the formula

$$\varepsilon = 1 + \frac{\langle M^2 \rangle - \langle M \rangle^2}{3\varepsilon_0 V k_B T} ;  \qquad \text{E 1}$$

and for the polarizable SWM4-DP model, similarly as in Lamoureux[6], using

$$\varepsilon = \varepsilon_\infty + \frac{\langle M^2 \rangle - \langle M \rangle^2}{3\varepsilon_0 V k_B T}, \quad \frac{\varepsilon_\infty - 1}{\varepsilon_\infty + 2} = \frac{4\pi\alpha}{3\langle v \rangle} ; \qquad \text{E 2}$$

where $M$ is the total dipole moment, $\varepsilon_0$ is the vacuum permittivity, $V$ is the volume of the system, $k_B$ is the Boltzmann-constant, T is the temperature, $<v>$ is the average molecular volume and $\alpha$ is the polarizability. For the polarizability for the SWM4-DP model the value of $\alpha$=1.04252 Å$^3$ from Lamoureux[21] et al. was used. The average molecular volume was $<v>=30.49478$ Å$^3$, yielding $\varepsilon_\infty=1.5$. All calculations applied the *g_dipoles* software provided by the GROMACS package[17]. Note that the exact equation by which the dielectric constant was calculated is not really important here (even though a large volume of the literature concerns this issue, see e.g.[22,23]). What we wish to scrutinize is the fluctuations of *M* (see E 1, E 2).

For the largest H4 system it was not possible to collect every consecutive configuration for 8 ns (due to the exceptionally large storage space requirement); for this reason, the effect of sampling only every *x*-th configuration was also studied. Results for each model are given in Table 2. There is no significant change if the dielectric constant is calculated based on every *x*=1-1000 configuration; on the other hand, *x*=10000 results in significantly different, sometimes lower, sometimes higher values. We can conclude that it is possible to use every 10-1000$^{th}$, or, to be on the safe side, every 10-100$^{th}$ configuration at least for this type of system, so that storage capacity and computation time can be optimized.

Convergence of the dielectric constant for the H3 model is given in Figure 3a. The curves for the *x*=1 and *x*=100 are identical for the entire period. The *x*=1000 curve joins them around t=1.4 ns, while the curve associated with the *x*=10000-th configuration sampling is distinctly different even in terms of the asymptotic value. This behavior is the same for each model in this work, which suggests that this is a general pattern.

Considering the convergence of the static dielectric constant it is clear that proper convergence can only be achieved after 3000-5000 ps. Usually much shorter time ranges have been used for the calculation of the static dielectric constant[6,15, 24,25,] which makes earlier results somewhat questionable. In Table 2 the average and the standard deviation calculated for the properly converged last 3000 ps of the trajectory are shown for $x=1$ curves ($x=100$ for H4).

Now let us turn to the system size dependence of the dielectric constant, as revealed here on the examples of the H1, H2, H3 and H4 models. Unfortunately, there is quite a large difference (68.3 –78.2) between the models, and no monotonous system size dependence can be established (Table 2, Figure 3b). This is similar to Morrow's findings[9], based on theoretical calculations for dipole lattices in case of reaction field, where the convergence was substantially oscillating with system size. Deviations between the different system sized models are much larger than the fluctuation of the static dielectric constant for any one model over the last 3000 ps of its trajectory, so they cannot be explained by improper convergence. To see how big a change in terms of the value of the dielectric constant can be found due to different starting points used in the calculation within the same model, $\varepsilon$ was calculated with $x=1$ from 200 ps to 8200 ps relative to the beginning of the data collection for the H1 model (see Figure 3c). Although there is a marked difference at the beginning of the simulations up to 4000 ps between this and the previous (0-8000 ps) calculations, later the dielectric constant would converge to the same value as before.

Comparing the 2000 molecule 293 K H3 and 298 K H3-T models the correct behavior, the decrease of the dielectric constant with increasing temperature, can be observed. The same can be seen for the 293 K H1 and 298 K H1-O systems, although here not only the temperature but also the simulation parameters differed slightly.

If we compare our $\bar{\varepsilon}$ with the value given for SPC/E water using a 500 ps trajectory of a 256 molecule system at 298 K by Svishchev[15], the static dielectric constant given by them is 71.0±6.1 for $H_2O$ and 68.5±6.6 for $D_2O$. The difference from our H1-O model's $\varepsilon$ value of 67.9±1.5 can be attributed to the different time step, system size and, most of all, the very different trajectory lengths. The values for the $D_2O$ models cannot be directly compared since the temperature in Svishchev's calculations was 298 K while in ours it was 293 K. Still, it has to be noted that no difference is expected due to the isotopic substitution, so the—slightly different— values found for the otherwise identical H1 (70.6±0.4) and D1 (71.5±0.53) models can only be attributed to numerical inaccuracies.

For the SWM4-DP model the original value of Lamoureux at 298 K was 79±5[6], while our 293 K simulation yielded 78.9±0.55, which can be explained by the different system size, temperature, time step, and slight differences in terms of the simulations.

### 5. Self-diffusion coefficient

The self-diffusion coefficient, *D*, was determined via two routes, calculating the mean-square displacement (MSD) from the coordinates and using the Einstein-relationship

$$\lim_{t \to \infty} \langle \|r(t) - r(0)\|^2 \rangle = 6Dt \qquad \text{E 3}$$

and by calculating the velocity auto-correlation function from the velocities, then integrating it according to the Green-Kubo formula

$$D = \frac{1}{3} \int_{0 \to \infty} \langle v(t) \cdot v(0) \rangle dt \ . \qquad \text{E 4}$$

The full trajectories used for the dielectric constant calculations were used for the MSD method, restarting the calculation at every 20 ps to improve statistics. As the MSD curves were appropriately linear, the MSD-*t* plot, without the first and last 10 %, was used for linear regression. The MSD was calculated by the *g_MSD* program of the GROMACS package[17]. Values calculated for the centre of mass of the molecules (as opposed to individual atoms) are given in Table 3. The effect of using every *x*-th configuration (*x*=10, 100, 1000, 10000) during the MSD calculation was also studied. As the MSD-*t* plots were linear, no significant change due to using different *x* values could be found in most of the cases.

The self-diffusion coefficient was also calculated from the velocity auto-correlation function, which was determined by the *g_velacc* program of the GROMACS package. Different cutoff times (in terms of the length of the velocity trajectory used) were applied during the calculation; the length of the VACF, often called the time lag, $t_{lag}$, was always the half of the actual length of the velocity trajectory, which varied from 4 to 1000 ps, depending on the system size, giving 2 to 500 ps long VACF-s. The whole trajectory for an auto-correlation calculation is $t_{traj}=N \cdot \Delta t$ long. Calculating the correlation function up to $t_{lag}=N/2 \cdot \Delta t$ meant the usage of the same *N/2* data points for each time difference to make the statistics balanced.

The convergence of the self-diffusion coefficient with time was investigated by changing the cutoff time of the integration and with this, the $t_{lag}$ of the VACF, for several models. *D-t* plots resulting from this procedure for the H1 model can be seen in Figure 4. Here the VACF was calculated for (a) three different 4 ps blocks with $t_{lag}$ = 2 ps; (b) six different 20 ps blocks of the trajectory giving $t_{lag}$=10 ps; (c) two 100 and (d) one 1000 ps blocks, giving (c) $t_{lag}$ =50 and (d) 500 ps, respectively. The last curve is truncated at 50 ps on the chart for sake of clarity. One of the $t_{lag}$ = 2 ps curve differs entirely from all the other curves, having a sharp minimum at *t*=1 ps, and the final value of *D*=1.7 cm$^2$/s, while the other two $t_{lag}$=2 ps calculation yielded D=1.86·10$^{-5}$ and 1.97·10$^{-5}$ cm$^2$/s. It is apparent that the $t_{lag}$ =10 ps *D-t* plots are oscillating quite strongly and remarkably differ from each other; for the six $t_{lag}$=10 ps calculations the final values range between 1.49 and 2.35·10$^{-5}$ cm$^2$/s. The $t_{lag}$=50 ps curves are less different from each other, but still there is sizable oscillation (the final values are D=1.99·10$^{-5}$ and 2.05·10$^{-5}$ cm$^2$/s). Only the $t_{lag}$=500 ps curve seems to be satisfactorily smooth (D=2.06·10$^{-5}$ cm$^2$/s). The self-diffusion coefficient can be determined more accurately via the following way: instead of calculating the integral and consequently, D, only once for the whole $t_0 \rightarrow t_{lag}$ interval giving a final D, it is better to calculate a running integral for $t_0 \rightarrow t$ , where 0<t≤$t_{lag}$, thus providing a D-t function. D values for the converged final part can be averaged, giving $\overline{D}$, and the error estimated as the standard deviation for the same range. The $\overline{D}$ calculated this way for the 2-10 ps interval ranges between 1.92(±0.3) and 2.2(±0.18)·10$^{-5}$ cm$^2$/s for the six different $t_{lag}$=10 ps calculations and $\overline{D}$ =2.00(±0.08)·10$^{-5}$ and 2.06(±0.2)·10$^{-5}$ cm$^2$/s for the $t_{lag}$=50 ps calculation. This demonstrates that shorter ranges may lead to a seriously wrong value of D, even if averaging is used. This system was the smallest among our models; still, its size, 216 water molecules, is in the range usually used for water calculations[1,6,15,26,27,28]. For larger systems the convergence is a bit better, even for $t_{lag}$=10 ps. The final and average D values for the longest $t_{lag}$ of the model and for $t_{lag}$ =10 and 50 ps are given in Table 3. Even in case of larger systems and longer $t_{lag}$ times, averaging is recommended for the properly converged parts. We must emphasize that the short, 1-2 ps, time lag usually used in the calculation of the self-diffusion coefficient gives rather arbitrary results, especially for smaller (216-518 molecules) systems which are still regularly used in MD calculations.

A linear behaviour of the long-time tail of the VACF against $t^{-3/2}$ is predicted by kinetic theory and found in hard sphere simulations[29]. This tail can be used to add an appropriate tail correction during the integration of D from the VACF[15]. As it can be seen in Figure 5, in case of the tail of the VACF calculated for shorter time lags ($t_{lag}$=2 ps, but even for 10 ps) it would be difficult to fit a straight line; only in case of the long, $t_{lag} \geq 50$ ps, calculation the linear behaviour is visible at $t^{-3/2} < 2$ ps$^{-3/2}$ ($t > 0.6$ ps). Note that here no tail correction is required. The effect of the system size is also visible: the $t_{lag} \geq 50$ ps curve is close to linear in the 0-2 ps$^{-3/2}$ range for the 518 molecule H2 model (Figure 5 b), while it is far from linear for the 216 molecule H1 model (Figure 5 a).

We consider the best estimates (written in bold in Table 3) of the self-diffusion coefficient those calculated from MSD using $x=10, 100, 1000$ and calculated from the longest time lag in case of VACF. It can be seen that for the larger system ($\geq 2000$ molecules), the values calculated via MSD and VACF are quite close, sometimes identical.

The self-diffusion constant of the deuterium containing D1 model is lower than the corresponding H1 model, as it is expected. The difference is much bigger in case of the VACF-based calculation than it is in case of MSD; seemingly, the isotopic substitution might affect the correctness of the two methods differently.

The dependence of $D$ on the system size was investigated in more detail. In accordance with previous findings[11,12,15,30], the value of the self-diffusion coefficient is monotonously increasing with increasing system size from 2.06 to 2.38·10$^{-5}$ cm$^2$/s for the VACF and from 1.97 to 2.39·10$^{-5}$ cm$^2$/s for the MSD calculations. According to Yeh[12], the relationship between the self-diffusion coefficient without the periodic boundary condition-caused finite size effect ($D_0$) and the $D_{PBC}$ calculated from a finite size model is the following for a cubic simulation cell with cell length of $L$:

| | |
|---|---|
| $D_{PBC} = D_0 - \dfrac{k_B T \xi}{6 \pi \eta L}$ | E 5 |

$\eta$ is the shear viscosity of the solvent and $\xi$ =2.837297 coming from the Ewald sum for a cubic lattice. In case the viscosity is not known, $D_0$ can be calculated from the interception of the $D_{PBC}$-$1/L$ plot with the *y*-axis (extrapolation to infinite system).

$D_0$ was determined for both the VACF and MSD calculated $D_{PBC}$ values of model H1, H2, H3 and H4, using equation E 5, then averaging the resulting $D_0$ values, and from the *y*-axis intercept as well (Figure 6). An $\eta$=9±0.07·10$^{-4}$ Pa·s shear viscosity value, estimated from SPC/E water molecular dynamics calculations of Smith[31] at 277 and 300 K and taking into account the experimental viscosity values, was used. The average $D_0$ resulting from equation E 5 are $\overline{D}_0$(VACF)=2.46·10$^{-5}$ cm$^2$/s and $\overline{D}_0$(MSD)=2.42·10$^{-5}$ cm$^2$/s, and from the *y*-axis intercept of the linear fit to the $D_{PBC}$-$1/L$ plot $D_0$(VACF) =2.50·10$^{-5}$ cm$^2$/s and $D_0$(MSD) =2.53·10$^{-5}$ cm$^2$/s. The discrepancy must be caused by the fact that there are some additional factors also involved in the system-size dependence of the self-diffusion coefficient, apart from the purely hydrodynamic consideration leading to E 5, as suggested by Yeh[12]. It is interesting that in our case for SPC/E water the values resulted from E 5 are lower than the ones coming from the linear fit extrapolated to infinite system size, although it was the other way around for the TIP3P water simulation of Yeh[12]. The experimental value is 1.89·10$^{-5}$ cm$^2$/s at 292.6 K, so all of the calculated values are too high.

We compared our results with the diffusion coefficient values of Spångberg[13] et al. for SPC(E) water in the NpT ensemble at 298 K. They used system sizes with 128, 256, 512 and 1024 water molecules compared to our system size investigation at 293 K in an NVT ensemble spanning 216, 518, 2000 and 16000 water molecules, so the results cannot easily be compared. Comparing the size-wise closest 512 molecules model of Spångberg with our 518 molecules models, their $D_{PBC}$(298 K)=2.57±0.03 10$^{-5}$ cm$^2$/s while our result is $D_{PBC}$=2.16± 0.11 10$^{-5}$ cm$^2$/s from MSD and $D_{PBC}$=2.12±0.16 10$^{-5}$ cm$^2$/s from VACF at 293 K, which can be accounted for by the different temperature and simulation technique. Comparing their 1024 molecules model's $D_{PBC}$=2.58±0.03 10$^{-5}$ cm$^2$/s with our 298 K H3-T model (2000 molecules) we found $D_{PBC}$=2.57± 0.14 10$^{-5}$ cm$^2$/s from MSD and $D_{PBC}$=2.60±0.07 10$^{-5}$ cm$^2$/s from VACF, which are in close agreement. These findings (based on, admittedly, rather small samples) suggest that applying the NpT ensemble results in systematically higher values of $D$ than they are from the NVT ensemble.

Comparing the result of the 216 molecule H1-O model ($D_{PBC}$=2.25·10$^{-5}$ cm$^2$/s from the VACF, 2.21·10$^{-5}$ cm$^2$/s from MSD) to the 2.24·10$^{-5}$ cm$^2$/s value reported by Svishchev[15] for a 256 molecule system calculated from a 1.5 ps time lag VACF, and considering the effect of different integration cutoffs and system sizes, these values are in good agreement. Correcting our values according to E 5 gives $D_0$=2.62·10$^{-5}$ cm$^2$/s from the VACF, 2.58·10$^{-5}$ cm$^2$/s from MSD and comparing it with the 2.3·10$^{-5}$ cm$^2$/s experimental value for 298.15 K we find that our values are again higher. Based on this it seems that the SPC/E model tends to overestimate the diffusion coefficient.

The best estimate of the diffusion constant for the SWM4-DP model is $D_{PBC}$=2.37·10$^{-5}$ cm$^2$/s from both MSD and VACF at 293 K, which is higher than obtained for the similar size H3 SPC/E model at the same temperature. Using the value $\eta$=8·10$^{-4}$ Pa·s for our 293 K model estimated from the $\eta$=8·10$^{-4}$ Pa·s shear viscosity at 298.15 K calculated by Lamoureux for the very similar SWM4-NDP model[32] (whose viscosity is lower than the experimental value) the corrected self-diffusion coefficient is $D_0$=2.56·10$^{-5}$ cm$^2$/s; this is much higher than the experimental 1.89·10$^{-5}$ cm$^2$/s.

## 6. Conclusions

(1) It was demonstrated that shorter than about 5(-6) ns trajectories are not applicable for providing a good estimate of the dielectric constant for any system size considered here; any value resulting from shorter calculations must be handled with care. It should be noted that, strictly speaking, calculation of the dielectric constant does not necessitate the presence of time dependence (dynamics); a statement that would be equivalent to the above could be that a volume of the configuration space that is comparable to the section sampled during about 5-6 ns in a molecular dynamics simulation should be used.

(2) The static dielectric constant changes strongly with the system size, too; however, the monotonous increase with the system size predicted by Morrow[9] could not be verified. Even long trajectory lengths cannot ensure an infinitely accurate estimation of the dielectric constant, due to the observed system size dependence: the difference between extrema is about 10 %.

(3) Concerning the calculation of the self-diffusion coefficient, the MSD method seems to be less sensitive to the circumstances of the calculation, even for smaller system sizes, provided that long enough trajectories are used. The VACF-based calculation is highly sensitive to the applied velocity trajectory length and consequently the cutoff time of the integration, even if averaging of the reasonably converged part of the *D-t* plot is applied. According to our findings the integral using a short time lag, of about 2 ps, can give a value quite different from the properly converged value using a much longer (50-500 ps) time lag; the extent of variations depends on the system size. Checking the convergence for the integral from the *D-t* plot visually is advisable and subsequently, averaging *D(t)* for the properly converged period is recommended. Even using the average of several short time lag VACF from several different parts of the trajectory is not as satisfactory as calculating the VACF for a longer time lag. Appropriate convergence can only be expected for much longer trajectory lengths and $t_{lag}$ than that had usually been used (for instance, $t_{lag}$=500 ps should be appropriate for a 216 water molecule system).

(4) The self-diffusion constant corrected to eliminate the finite-size effect of SPC/E water at 293 K was found to be higher than the experimental value. For the SWM4-DP model both the MSD and the longest $t_{lag}$ VACF calculation yielded even higher values of the self-diffusion constant.

(5) According to our findings, the system size, a small temperature difference and the applied trajectory length caused hardly any change in terms of the radial distribution functions and the static structure factor. The difference in the potential model resulted, on the other hand, in significant differences when the SPC/E and the SWM4-DP models were compared.

(6) It is therefore clearly demonstrated by the present study that comparing the accuracy/goodness of potential models based on their ability to reproduce the experimental static dielectric constant and self-diffusion coefficient (and other similar properties), as it is often done in the literature, can be misleading. In most cases, too short trajectories are used and in case of the self diffusion coefficient the values are rarely corrected for finite-size effects. We suggest that structural properties such as *g(r)* and *S(Q)* should receive greater attention during the validation of an interaction potential, as these are less sensitive to the model size and calculation method.

**Table 1** Basic parameters of the simulation models: name, chemical composition, number of molecules ($N$), temperature, potential model, time step ($dt$), data collection length for the coordinates and velocities.

| name | system | $N$ | $T$ (K) | potential | $dt$ (fs) | length for coord. (ns) | length for vel. (ps) |
|---|---|---|---|---|---|---|---|
| H1 | $H_2O$ | 216 | 293 | SPC/E | 2 | 8.2 | 1000 |
| H2 | $H_2O$ | 518 | 293 | SPC/E | 2 | 8 | 400 |
| H3 | $H_2O$ | 2000 | 293 | SPC/E | 2 | 8 | 200 |
| H4 | $H_2O$ | 16000 | 293 | SPC/E | 2 | 8 | 100 |
| H1-O | $H_2O$ | 216 | 298 | SPC/E | 1.25 | 6 | 625 |
| D1 | $D_2O$ | 216 | 293 | SPC/E | 2 | 8 | 1000 |
| H3-T | H2O | 2000 | 298 | SPC/E | 2 | 8 | 200 |
| SWM4-DP | H2O | 2000 | 293 | SWM4-DP | 2 | 8 | 200 |

**Table 2** The static dielectric constant calculated from every $x$th configuration of the trajectory. '$\varepsilon$ final' is the final value, $\overline{\varepsilon}$ is the average and $\sigma$ is the standard deviation calculated for the properly converged, last 3000 ps part of the x=1 trajectory (*: the average for the H4 system is calculated from the x=10 curve).

| Model | $\varepsilon$ final | | | | | $\overline{\varepsilon}$, ($\sigma$) |
|---|---|---|---|---|---|---|
| | $x=1$ | $x=10$ | $x=100$ | $x=1000$ | $x=10000$ | $x=1$ * |
| H1 | 69.8 | - | 69.8 | 69.7 | 68.1 | 70.6 (0.40) |
| H2 | 65.6 | - | 65.7 | 66.5 | 66.5 | 65.4 (0.44) |
| H3 | 77.9 | - | 77.9 | 77.9 | 79.4 | 78.2 (0.40) |
| H4 | - | 72.4 | 72.4 | 72.3 | 69.7 | 72.4 (0.66) |
| H1-O | 67.1 | - | 67.1 | 66.9 | 66.0 | 67.9 (0.71) |
| D1 | 71.5 | - | 71.5 | 70.9 | 69.8 | 71.5 (0.53) |
| H3-T | 72.3 | - | 72.3 | 72.5 | 68.4 | 71.5 (0.59) |
| SWM4-DP | 78.1 | - | 78.2 | 77.4 | 78.8 | 78.9 (0.55) |
| exp (293 K) | | | | | | 80.4[24] |
| exp (298 K) | | | | | | 78.4[25] |

**Table 3** The self-diffusion coefficient of the models calculated either from MSD or from VACF. In case of MSD, values using every $x=10$, 100 or 1000th configuration for sampling are the same. In case of VACF, the final time value, the so-called 'lag' of the VACF is given as $t_{lag}$ and two $D$ values are provided for each $t_{lag}$ the final value of the integral (which in case of more than one calculation for the given $t_{lag}$ is averaged). 'avr.' means the average $D$, calculated from the $D$-$t$ integral in the $2 - t_{lag}$ ps range, (also averaged in case of more than one calculation for the given $t_{lag}$)The longest $t_{lag}$ is different for the different models: $Y=312.5$ ps for H1-O and 500 ps for the other 216 molecule system; 200 ps for the 518, 100 ps for the 2000 molecule models. For H4 the longest $t_{lag}$ was 50 ps.

| Model | $D_{MSD}$ ($10^{-5}$ cm$^2$/s) | | $D_{VACF}$ ($10^{-5}$ cm$^2$/s) | | | | | |
|---|---|---|---|---|---|---|---|---|
| | $x=10,100,1000$ | $x=10000$ | ($t_{lag}=10$ps) | | ($t_{lag}=50$ps) | | ($t_{lag}=Y$ ps) | |
| | | | final | avr. | final | avr. | final | avr. |
| H1 | **1.97 (0.12)** | 1.89 (0.02) | 2.03 (0.14) | 2.05 (0.12) | 2.03 (0.2) | 2.03 (0.03) | 1.98 | **2.06 (0.04)** |
| H2 | **2.16 (0.11)** | 2.18 (0.10) | 2.21 | 2.20 (0.14) | 2.27 | 2.24 (011) | 2.38 | **2.21 (0.16)** |
| H3 | **2.28 (0.01)** | 2.28 (0.00) | 2.23 | 2.33 (0.08) | 2.38 | 2.32 (0.07) | 2.32 | **2.31 (0.06)** |
| H4 | **2.39 (0.02)** | 2.39 (0 02) | 2.43 | 2.44 (0.02) | 2.35 | **2.38 (0.06)** | | |
| H1-O | **2.21 (0.14)** | 2.22 (0.09) | 2.10 | 2.17 (0.16) | 2.22 (0.17) | 2.33 (0.08) | 2.30 | **2.25 (0.13)** |
| D1 | **1.94 (0.29)** | 1.86 (0.46) | 1.36 | 1.57 (0.21) | 1.66 | 1.74 (0.12) | 1.9 | **1.85 (0.11)** |
| H3-T | **2.57 (0.14)** | 2.57 (0.14) | 2.66 | 2.71 (0.07) | 2.54 | 2.59 (0.09) | 2.66 | **2.60 (0.07)** |
| SWM4-DP | **2.37 (0.00)** | 2.37 (0.00) | 2.47 | 2.48 (0.08) | 2.38 | 2.32 (0.15) | 2.30 | **2.37 (0.11)** |
| exp (H$_2$O 292.6 K) | 1.89[26] | | | | | | | |
| exp ( H$_2$O, 298.15 K) | 2.3[26] | | | | | | | |

**Figure captions**

**Figure 1:** Partial radial distribution functions for the various models. (a) The effect of system size for the 293 K H1, H2, H3 and H4 model. (b) The effect of temperature and slightly different run parameters (H1, H1-O) and the isotope effect (D1) for the 216 molecules models. The O-H (O-D) partials are plotted against the left axis with no shift and the O-O partials shifted by 4 on the same axis, the H-H (D-D) partials are plotted against the right axis for the sake of clarity.

**Figure 2:** (a) Radial distribution functions and (b) the (neutron weighted) total scattering static structure factor for the H3 and SWM4-DP models.

**Figure 3:** Convergence of the static dielectric constant for (a) H3, (b) H1, H2, H3, H4; $x=100$ for all four systems, (c) H1 model. Every $x$th configuration was used for the calculation of the running averages; $x$ is given on the graph if not stated explicitly here. In case of (b) a second calculation range with $x=1$ from 200-8200 ps is also shown, which is indicated by an arrow for clarity.

**Figure 4:** Convergence of the self-diffusion coefficient calculated from the VACF for the H1 model. The time intervals of the trajectory ($2 \cdot t_{lag}$) for which the $D$-$t$ curves were calculated are given on the chart.

**Figure 5:** The tail of the VACF for the (a) H1 and (b) H2 models. The time lag values identify the curves; in some of the cases the part of the trajectory for which they were calculated are given, too.

**Figure 6:** Finite-size effect for the VACF and MSD-calculated self-diffusion coefficient values of the H1, H2, H3 and H4 models. $D_{PBC}$ values for both calculation methods with the linear fit to them are shown, and the corrected $D_0$ values with the averaged $D_0$ shown by dotted lines.

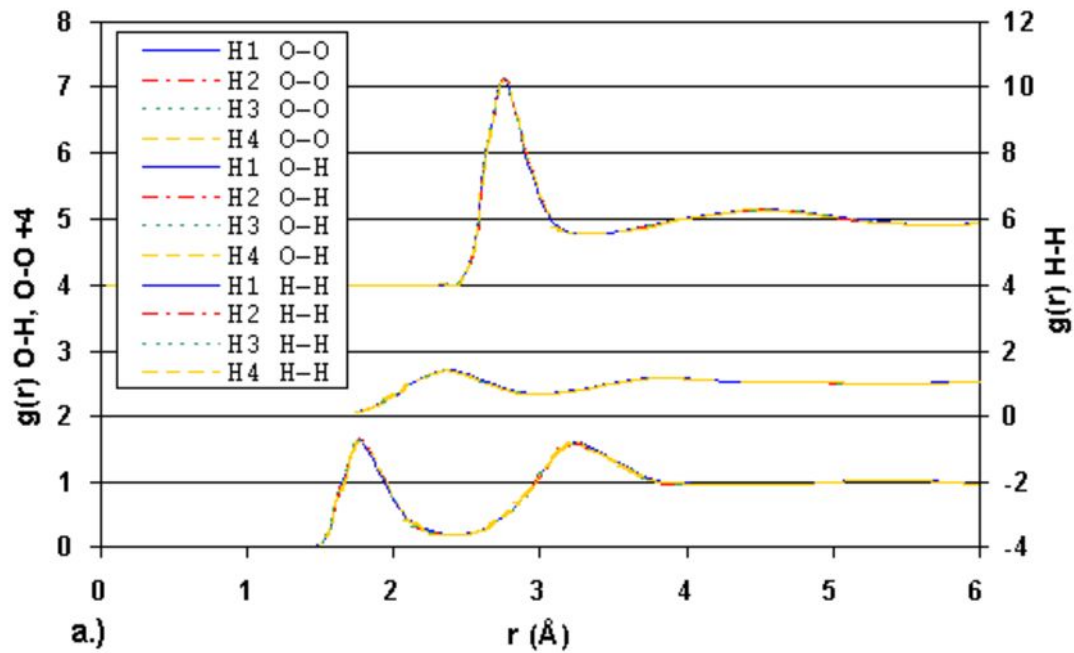

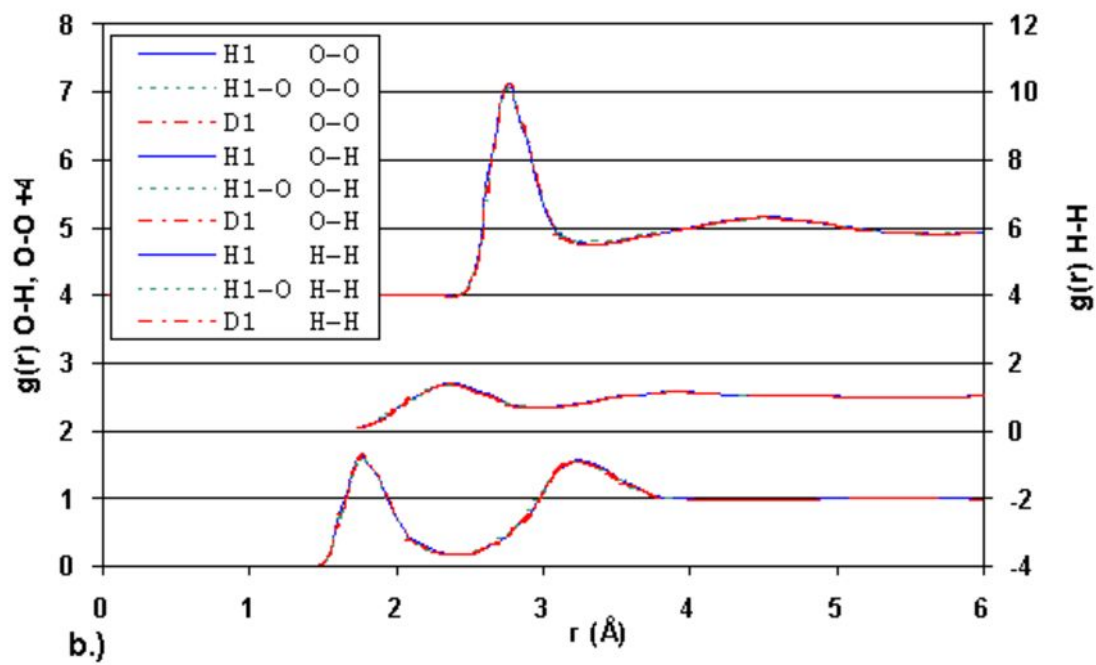

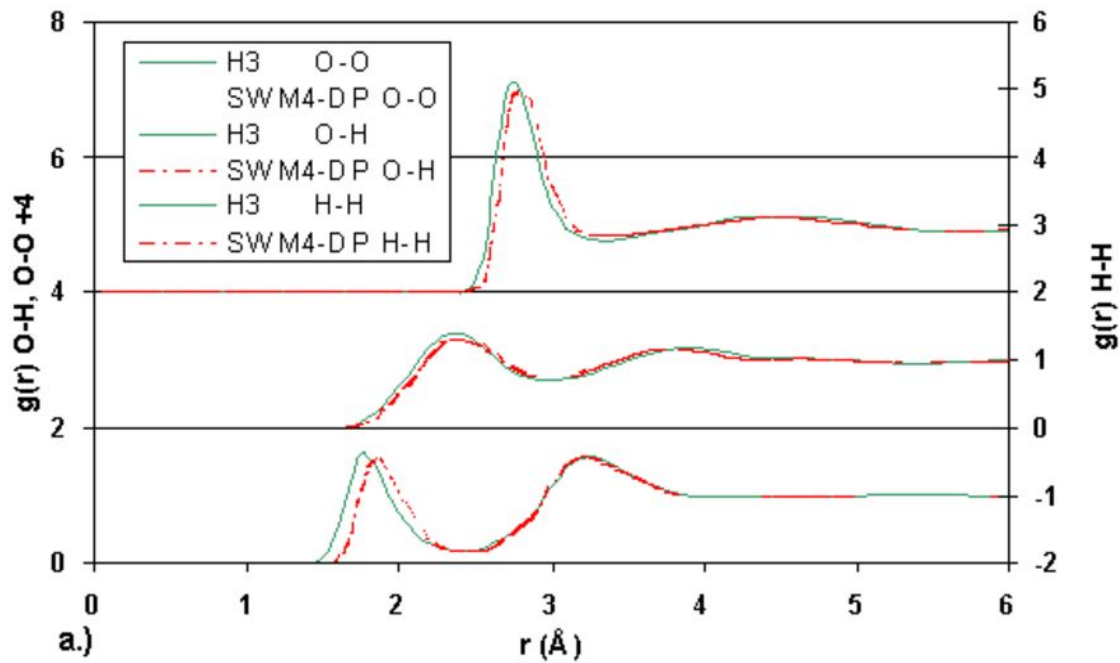

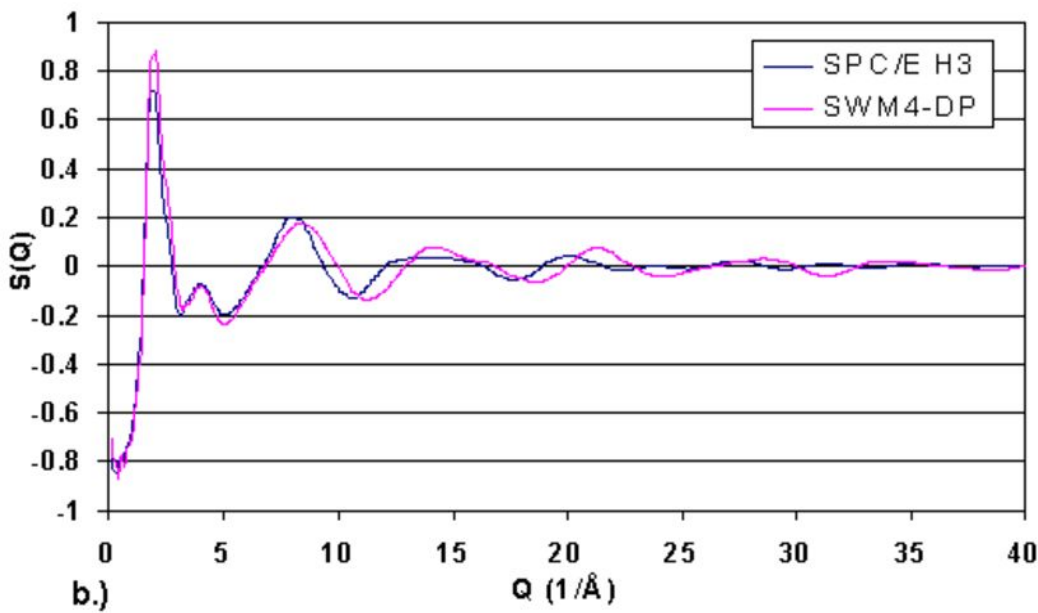

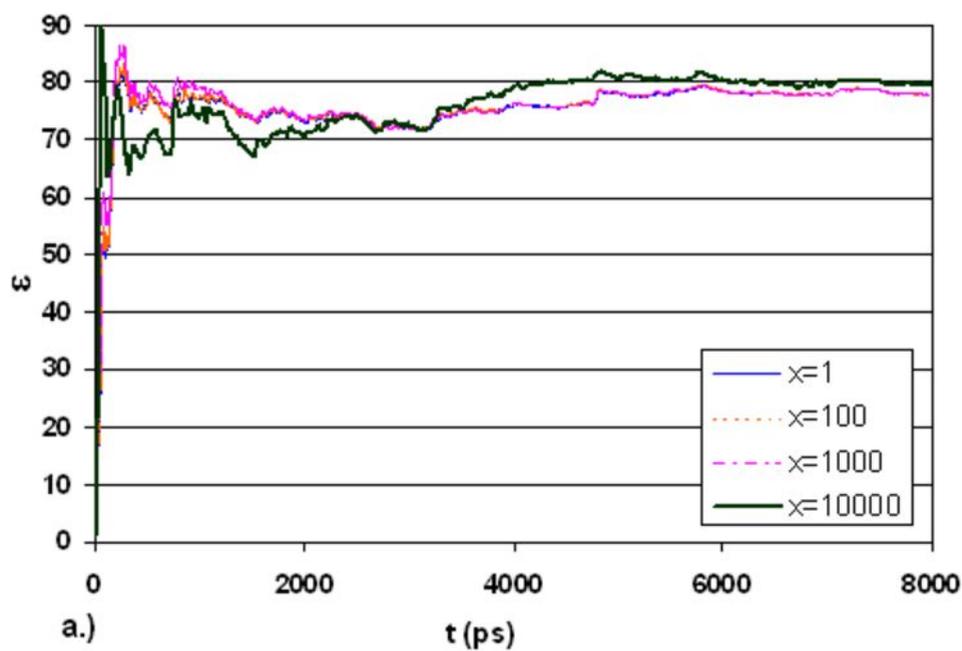
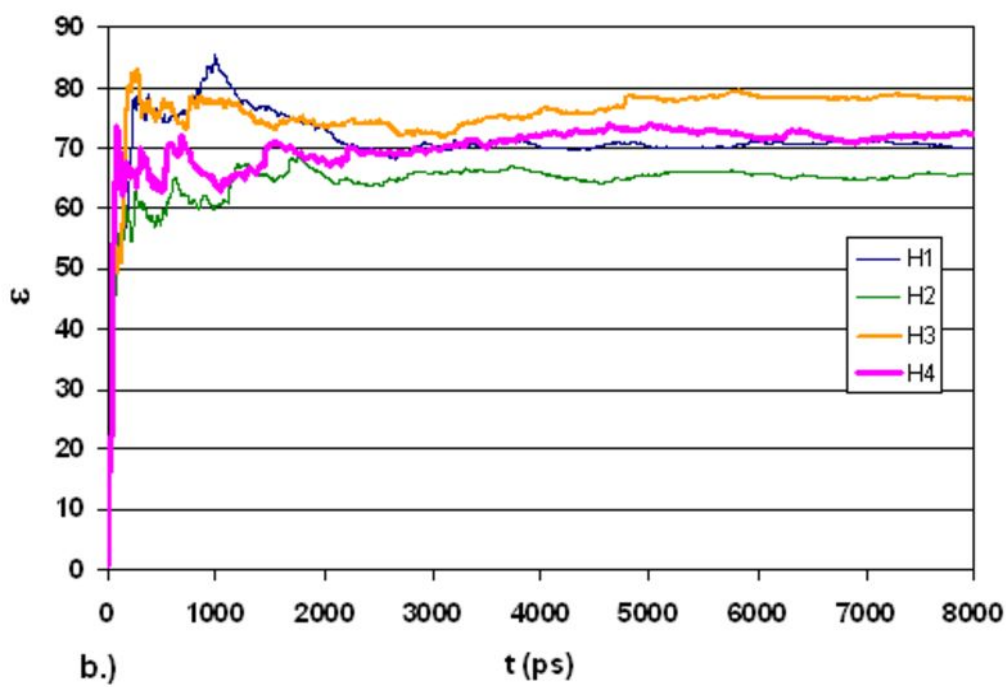
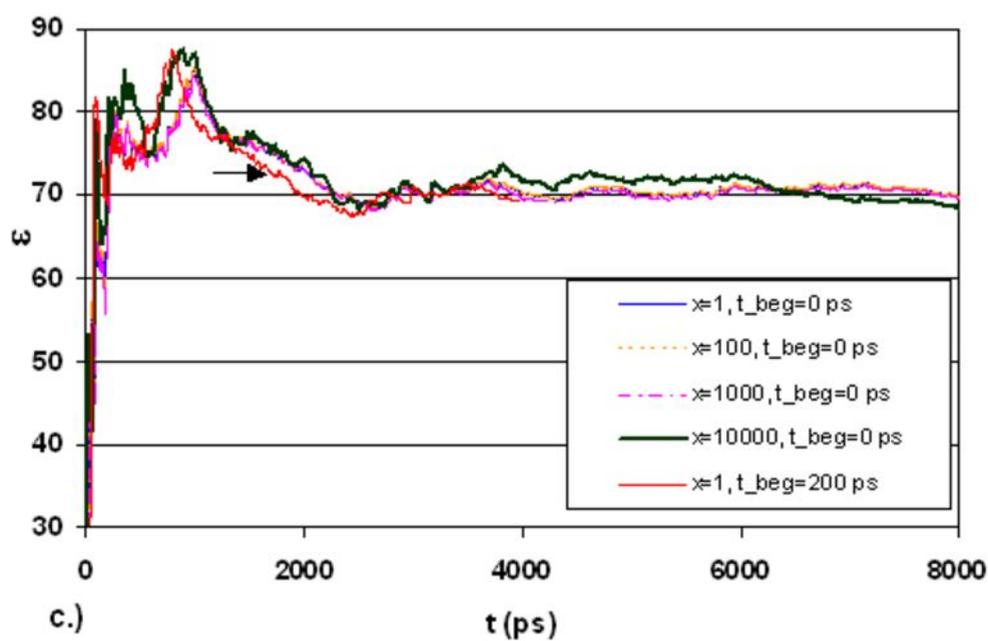

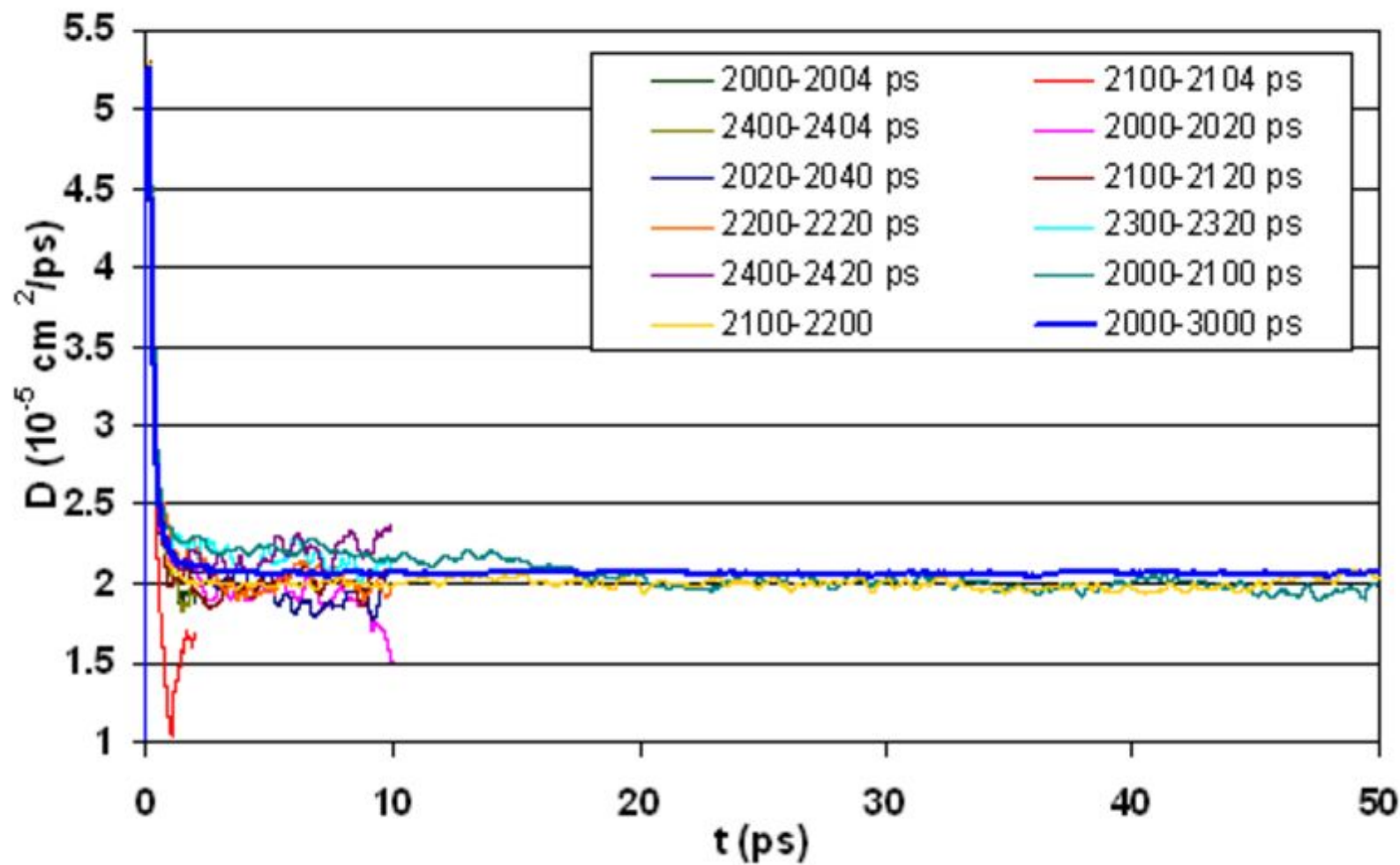

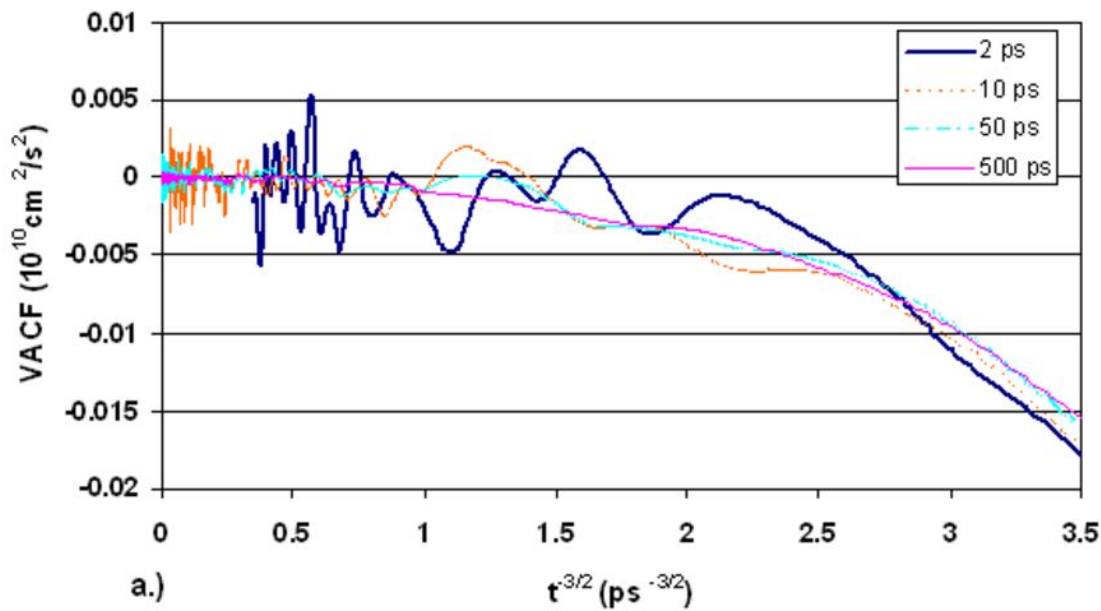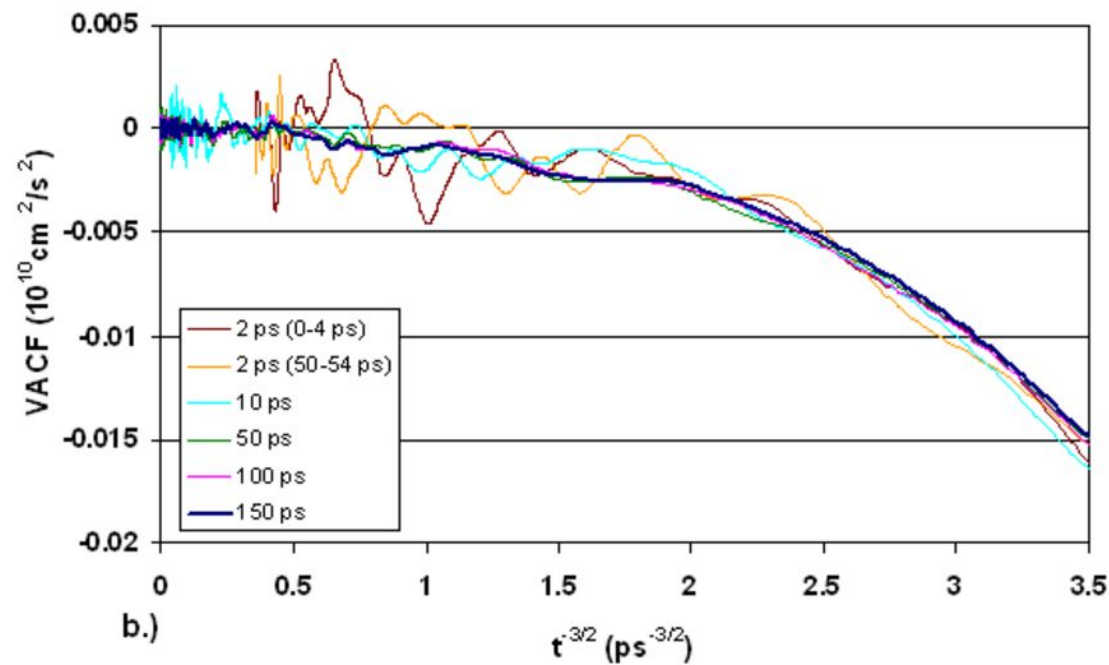

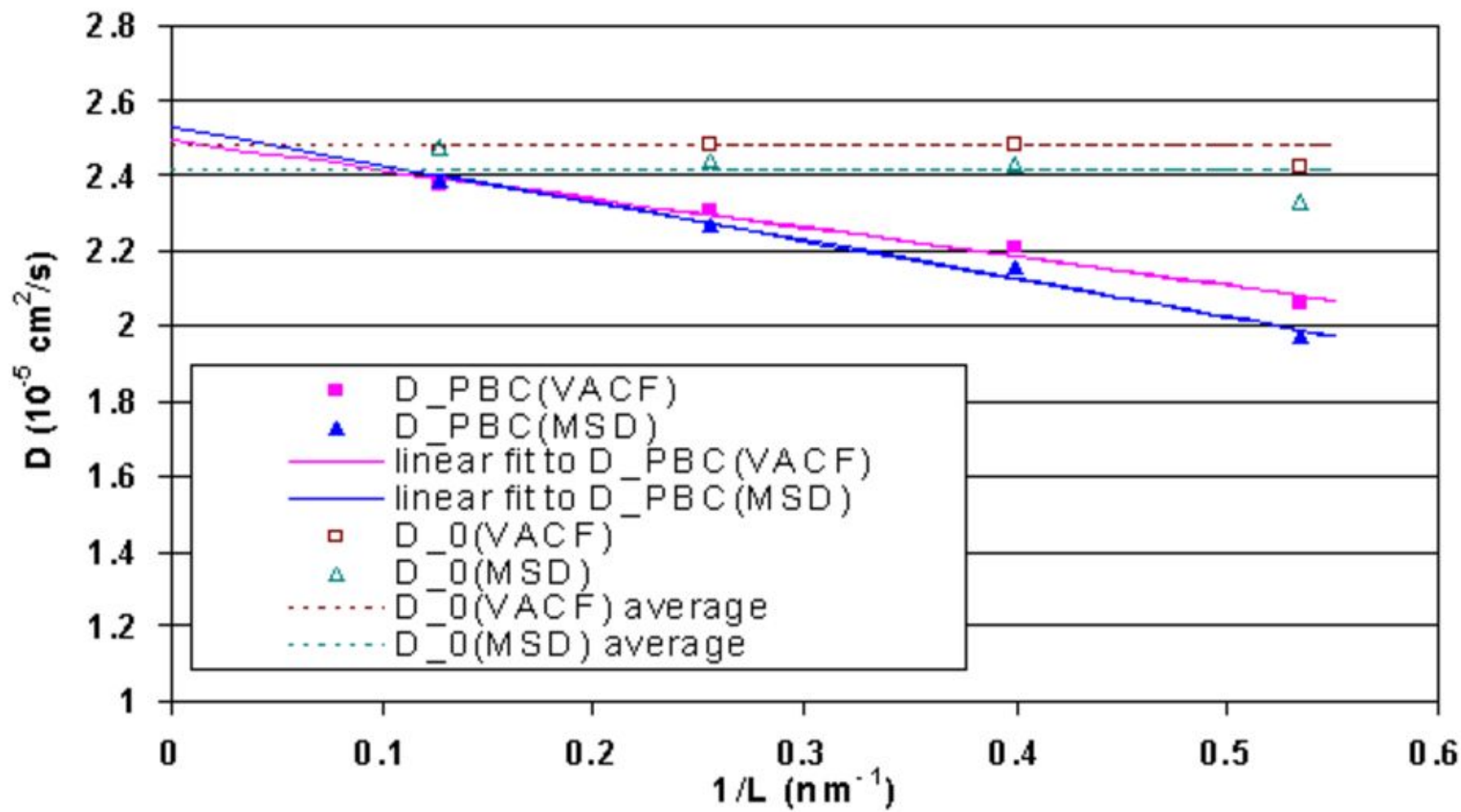